\documentclass[preprint,showpacs,preprintnumbers,amsmath,amssymb]{revtex4}

\usepackage{epsfig}
\usepackage{hyperref}

\begin{document}
\title{$B_{(s)}\to S$ transitions in the light cone sum rules
with the chiral current}
\author{Yan-Jun Sun$^{1,\;2}$\footnote{Email: sunyj@ihep.ac.cn},
 Zuo-Hong Li$^3$\footnote{Corresponding author~~Email: lizh@ytu.edu.cn} and Tao Huang$^2$
\footnote{Corresponding author~~ Email: huangtao@ihep.ac.cn}}
\address{$^1$Department of Modern Physics, University of Science and
Technology of China, Hefei 230026, P.R. China\\
$^2$Institute of High Energy Physics and Theoretical Physics Enter
for Science Facilities,Chinese Academy of Sciences,
Beijing 100049, P.R. China\\
$^3$Department of Physics, Yantai University, Yantai 264005,China}

\begin{abstract}
We make a QCD light-cone sum rule (LCSR) assessment of $B_{(s)}$
semi-leptonic decays to a light scalar meson, $B_{(s)}\to S
l\bar{\nu}_l, S l \bar{l}\,\,(l=e,\mu,\tau)$. Chiral current
correlators are used and calculations are performed at leading order
in $\alpha_s$. Having little knowledge of ingredients of the scalar
mesons, we confine ourself to the two quark picture for them and
work with the two possible scenarios. The resulting sum rules for
the form factors receive no contributions from the twist-3
distribution amplitudes (DA's), in comparison with the calculation
of the conventional LCSR approach where the twist-3 parts play
usually an important role. We specify the range of the squared
momentum transfer $q^2$, in which the operator product expansion
(OPE) for the correlators remains valid approximately. It is found
that the form factors satisfy a relation consistent with the
prediction of soft collinear effective theory (SCET). In the
effective range we investigate behaviors of the form factors and
differential decay widthes and compare our calculations with the
observations from other approaches.

\end{abstract}


\maketitle

\section{introduction}

With numerous scalar meson states being discovered experimentally,
most of efforts have been devoted to a study about their inner
structure and how they are classified. However, much controversy
 persists regarding the underlying components
of them. Currently one of our main concerns is that these scalar
particles can whether or not be described consistently in a quark
picture. Recently, from a survey of the accumulated experimental
data the two possible scenarios are suggested
\cite{Cheng:2005nb,Lu:2006fr}, where the scalar mesons below and
above $1 ~\mbox{GeV}$ are assumed to enter their respective nonets
in two different ways. In scenario 1, there are the two scalar
nonets formed by the two quark bound states. One contains, as the
lowest lying scalar states, the isoscalars $\sigma(600)$ and
$f_0(980)$, isodoublets $(\kappa^+(800), \kappa^0(800))$ and
$(\bar{\kappa}^0(800), \kappa^-(800))$ and isovector $(a_0^+(980),
a_0^0(980), a_0^-(980))$. The other is made up of the corresponding
first excited states: the isoscalars $f(1370)$ and $f_0(1500)$,
isodoublets $(K^{*+}_0(1430), K^{*0}_0(1430))$ and
$(\bar{K}^{*0}_0(1430), \bar{K}^{*-}_0(1430))$ and isovector
$(a_0^+(1450), a_0^0(1450), a_0^-(1450))$. In scenario 2, those
scalar states below $1 ~\mbox{GeV}$ are taken to be the members of a
four-quark nonet, while $f(1370)$, $f_0(1500)$, $a_0(1450)$ and
$K^*_0(1430)$ are treated as the lowest lying two-quark resonances
and arranged into another nonet, with the corresponding first
excited states between $2.0\sim2.3~\mbox{GeV}$.

Although now we are not able to discriminate among all the existing
schemes for the scalar mesons, the above two are intriguing in that
they can provide us with a ground to make a systematic study on the
scalar mesons. In such assignment scenarios, an investigation has
been made into the related decay constants and light-cone
distribution amplitudes (DA's)\cite{Cheng:2005nb}. More importantly,
to gain insight into the scalar mesons some of the B decays
involving them have been explored in the same context. In
Ref.\cite{Cheng:2005nb,Lu:2006fr}, the hadronic decays with a scalar
final state are discussed in detail in the framework of QCD
factorization, important implications being drawn for the properties
of the scalar particles. More attentions are paid to the
semileptonic decays with a potential interest $B_{(s)}\to S l
\bar{\nu}_l, S l \bar{l}$. Especially, one shows a great interest in
the knowledge of their differential rates, since it is critical, as
confronted with the coming experimental observations, for acquiring
valuable information on ingredients of the scalar particles.
Unfortunately, among the existing approaches no one can afford the
task to understand the underlying form factors in the whole regions
of $q^2$, with $q$ being the momentum transfers. An effective range
of $q^2$, in which the calculations are believable, has even not
been specified in literature, the computations being carried out in
just a small or intermediate kinematical region arbitrarily
selected. So the results are less persuasive.

Superior to the three-point QCD sum rules in evaluating heavy to
light meson transitions, the LCSR approach, which starts with a
two-point correlation function, adopts the operator product
expansion (OPE) near the light cone $x^2=0$ in terms of nonlocal
operators, whose matrix elements are parameterized as the hadronic
DA's of increasing twist. Such that the resulting LCSR for form
factors, in addition to having an estimable effective region of
$q^2$, can embody as many long-distance effects as possible involved
in the decaying processes. However, a better understanding of these
DA's is critical to have the calculation more reliable. Together
with the leading twist-2 DA, in general, the twist-3 ones enter and
play an important role in a LCSR calculation on the form factors. In
the case of the scalar mesons, the probe into the twist-2 and -3
DA's has been conducted in the framework of QCD sum rules and a DA
model, in an expansion form in the Gegenbauer polynomials, has been
formulated, but with a sizable error in some of the model
parameters. To try our best to reduce uncertainty in LCSR
calculation from the long distance parameters, a practical
improvement scenario has been worked out with its validity examined
and confirmed, in which a chiral correlator is so chosen that the
twist-3 DA's make no contribution \cite{Huang:2001xb}. In the
present work, we intend to apply the same trick to revaluate the
semileptonic transitions $B_{(s)}\to S l\bar{\nu}_l, S l \bar{l}$,
in the two quark picture for the scalar mesons. We will work in the
effective regions required by the OPE validity and with the two
different scenarios aforementioned, and calculation is to be
performed at leading order in $\alpha_s$.

The paper is organized as follows: In the following section, we
present the correlation functions with a chiral current and use them
to derive the LCSR for the form factors for the $B_{(s)} \to S $
transitions. The discussion and comment are made on the important
inputs-the DA's and decay constants of the scalar mesons, in
Sec.III. Sec.VI is devoted to a detailed numerical discussion about
the form factors and differential widths for $B_{(s)}\to S l
\bar{\nu}_l, S l \bar{l}$, including a numerical comparison with the
estimates of some other approaches. The final section is reserved
for a summary.

\section{The LCSR for the $B_{(s)}\to S$ form factors}
 \begin{table}[b]
 \caption{The values of Wilson coefficients $C_i(m_b)$ in the leading
logarithmic approximation in Standard Model, with
$m_W=80.4\mbox{GeV}$, $m_t=173.8\mbox{GeV}$,
$m_b=4.8\mbox{GeV}$\protect\cite{Hatanaka:2008gu}.}
 \label{tab:wilsons}
 \begin{center}
 \begin{tabular}{c c c c c c c c c}
 \hline\hline
 \ \ \ $C_1$ &$C_2$ &$C_3$ &$C_4$ &$C_5$ &$C_6$ &$C_7$ &$C_9$ &$C_{10}$       \\
 \ \ \ $1.119$   &$-0.270$   &$0.013$    &$-0.027$    &$0.009$    &$-0.033$    &$-0.322$    &$4.344$    &$-4.669$    \\
 \hline\hline
 \end{tabular}
 \end{center}
 \end{table}

In the standard model (SM), the semileptonic decays $B_{(s)}\to S l
\bar{\nu}_l, S l \bar{l}$ are induced by the following effective
Hamiltonian:
\begin{eqnarray}
{\cal H}_{eff}&=&\frac{G_F}{\sqrt{2}}V_{ub}\bar
 u\gamma_{\mu}(1-\gamma_5)b \bar{\ell}\gamma^{\mu}(1-\gamma_5)\nu_{\ell}\nonumber \\
&+&\frac{G_F\alpha V_{tb}^*V_{ts}}{\sqrt{2}\pi }\Big[ C_9^{\rm{eff}}
\bar{s}\gamma _{\mu }(1-\gamma_5)b\ \bar{\ell}
       \gamma ^{\mu }\ell\nonumber \\
      &+&C_{10}\bar{s}\gamma _\mu (1-\gamma_5)b
       \bar{\ell}\gamma ^{\mu }\gamma _5 \ell\nonumber \\
    &-&\frac{2m_b C_7^{\rm{eff}}(m_b)}{q^2}
       \bar{s} i\sigma _{\mu \nu }
       q^{\nu }(1+\gamma_5)b\ \bar{\ell}\gamma ^{\mu }\ell\Big].
       \label{Hll}
 \end{eqnarray}
Here $V_{ij}$ are the Cabibbo-Kobayashi-Maskawa (CKM) matrix
elements, and $C_{i}^{\rm{(eff)}}$ the Wilson coefficients, among
which $C_{9}^{\rm{eff}}$ and $C_{10}$ are scale-independent for the
corresponding operators have a vanishing anomalous dimension.
$C_7^{\rm{eff}}$ and $C_9^{\rm{eff}}$ are expressed as
 \begin{eqnarray}
C_7^{\rm{eff}}(\mu)&=&C_7(\mu)+C_{b\to s\gamma}(\mu),\\
 C_9^{\rm{eff}}&=&C_9(\mu)+Y_{\rm{pert}}(s')+Y_{\rm{LD}}(s'),
 \end{eqnarray}
where $C_{b\to s\gamma}(\mu)$ stems from the absorptive part of
$b\to s c\bar c\to s\gamma$ rescattering which will be neglected
here, $Y_{\rm{pert}}$ and $Y_{\rm{LD}}$ stand for, respectively, the
short-and long-distance contributions from the four quark operators
\cite{Buras:1994dj}, with
 \begin{eqnarray}
 Y_{\rm{pert}}(s')&=&
 h(z,s')C_0-\frac{1}{2}h(1,s')(4C_3+4
 C_4+3C_5+C_6)\nonumber\\
 &&-\frac{1}{2}h(0,s')(C_3+3
 C_4) + \frac{2}{9}(3C_3 + C_4 +3C_5+ C_6),\label{eq:ypert}
 \end{eqnarray}
$C_0=3C_1+C_2+3C_3+C_4+3C_5 +C_6$, and
\begin{eqnarray}
h(z,s') &=& -{8\over 9}{\rm {ln}}z+{8\over 27}+{4\over 9}x-{2\over
9}(2+x)|1-x|^{1/2}
 \left\{
\begin{array}{l}
\ln \left| \frac{\sqrt{1-x}+1}{\sqrt{1-x}-1}\right| -i\pi \quad {\rm {for}}%
{ {\ }x\equiv 4z^{2}/s^{\prime }<1} \\
2\arctan \frac{1}{\sqrt{x-1}}\qquad {\rm {for}}{ {\ }x\equiv
4z^{2}/s^{\prime }>1}
\end{array}
\right., \nonumber\\
h(0,s^{\prime}) &=& {8 \over 27}-{8 \over 9} {\rm ln}{m_b \over \mu}
-{4 \over 9} {\rm {ln}}s^{\prime} +{4 \over 9}i \pi \,\, ,
\end{eqnarray}
where $z=m_c/m_b$ and $s^{\prime}=q^2/m^2_b$. The Wilson
coefficients $C_i(m_b)$, listed in Tab.\ref{tab:wilsons}, are given
in the leading logarithmic accuracy.

Aiming at an evaluation of the semileptonic decays $B_{(s)}\to S l
\bar{\nu}_l, S l \bar{l}$, we need to confront the hadronic matrix
elements $\langle S(p)|\bar {q_2}\gamma_{\mu}\gamma_5b|
B_{(s)}(p+q)\rangle$ and $\langle S(p)|\bar
{q_2}\sigma_{\mu\nu}\gamma_5q^\nu b| B_{(s)}(p+q)\rangle$. They can
be parameterized, in terms of the form factors $f_+(q^2)$,
$f_-(q^2)$ and $f_T(q^2)$, as
\begin{eqnarray}
 \langle S(p)|\bar q_2\gamma_{\mu}\gamma_5b| B(p+q)\rangle
  &=& -2ip_\mu
  f_{+}(q^2)-i[f_{+}(q^2)+f_{-}(q^2)]q_\mu,\label{eq:BS1}\\
   \langle S(p)|\bar q_2\sigma_{\mu\nu} \gamma_5q^\nu b|  B(p+q)\rangle
  &=& [2p_\mu q^2-2q_\mu (q\cdot p)]\frac{-f_T(q^2)}{m_B+m_S},
  \label{eq:BS2}
\end{eqnarray}
where the $B_{(s)}$ mesons are signified by $B$ for short. The
relative form factors could be calculated in the LCSR. Instead of
the correlation functions used in Ref.\cite{Wang:2008da}, we would
like to consider the following two correlators, with the T product
of chiral current operators sandwiched between the vacuum and one
on-shell scalar meson state \cite{Chernyak:1990ag}:
\begin{eqnarray}
\Pi_{\mu}(p,q)&=& i \int d^4 xe^{iqx} \langle S(p)|T\{\
\bar{q_2}(x)\gamma_{\mu}(1-\gamma_5)b(x),\bar{b}(0)i(1-\gamma_5)q_1(0)\}
|0 \rangle, \label{eq:pis1}\\
\widetilde{\Pi}_{\mu}(p,q)&=& i \int d^4 xe^{iqx} \langle S(p)|T\{\
\bar{q_2}(x)\sigma_{\mu\nu}(1+\gamma_5)q^{\nu}b(x),\bar{b}(0)i(1-\gamma_5)q_1(0)\}
|0 \rangle, \label{eq:pis2}
\end{eqnarray}
where $q_1,q_2$ denotes the light quark field.

The hadronic representations for them are easy to achieve, by
inserting between the currents a complete set of resonance states
with the same quantum numbers as the operator
$\bar{b}(0)i(1-\gamma_5)q_1(0)$. On the desired pole contributions
due to the lowest pseudoscalar $B$-meson are insolated, we obtain
the hadronic representations:
\begin{eqnarray}
\Pi_{\mu}^h (p,q)&=&\frac{\langle S(p)|\bar
{q}_2\gamma_{\mu}\gamma_5b|B(p+q)\rangle \langle
B(p+q)|\bar{b}i \gamma_5 q_1 |0\rangle}{m^2_B-(p+q)^2}\nonumber\\
&&+\sum_h \frac{\langle S(p)|\bar
{q}_2\gamma_{\mu}(1-\gamma_5)b|B^h(p+q)\rangle \langle
B^h(p+q)|\bar{b}i(1-\gamma_5) q_1
|0\rangle}{m^2_B-(p+q)^2},\label{eq:sp1}\\
\widetilde{\Pi}_{\mu}^h (p,q)&=&-\frac{\langle S(p)|\bar
{q}_2\sigma_{\mu\nu}(1+\gamma_5)q^\nu b|B(p+q)\rangle \langle
B(p+q)|\bar{b}i \gamma_5 q_1 |0\rangle}{m^2_B-(p+q)^2}\nonumber\\
&&+\sum_h \frac{\langle S(p)|\bar
{q}_2\sigma_{\mu\nu}(1+\gamma_5)q^\nu b|B^h(p+q)\rangle \langle
B^h(p+q)|\bar{b}i(1-\gamma_5)
q_1|0\rangle}{m^2_B-(p+q)^2}.\label{eq:sp2}
\end{eqnarray}
It should be stressed that the correlation functions receive
contributions from the scalar resonances included in the
intermediate states$B^h$\cite{Chernyak:1990ag}, in addition to the
higher pseudoscalar ones, and the ground-state scalar meson is a bit
lighter than the pseudoscalar resonance lying in the first excited
state.

With the definitions of $B$-meson decay constant $\langle B|\bar b
i\gamma_5 q_1 |0 \rangle=\frac{m_B^2 f_B}{m_{q_1}+m_b}$ and
Eqs.(\ref{eq:BS1}) and (\ref{eq:BS2}), the phenomenological
representations of the correlation functions read
\begin{eqnarray}
\Pi_{\mu}^h (p,q)&=&
\frac{-i}{m^2_B-(p+q)^2}\frac{m^2_Bf_B}{m_{q_1}+m_b}
 [2 f_{+}(q^2)p_\mu+(f_{+}(q^2)+f_{-}(q^2))q_\mu]\nonumber\\
 &&-\frac{1}{\pi}\int_{s_0}^{\infty} ds
\frac{2\rho^h_+(s)p_\mu+(\rho^h_+(s)+\rho^h_-(s))q_\mu}{s-(p+q)^2},\label{eq:spec1}\\
\widetilde{\Pi}_{\mu}^h (p,q)
 &=&\frac{1}{m^2_B-(p+q)^2}\frac{m^2_B f_B}{m_{q_1}+m_b}
\frac{f_T}{m_B+m_S}[2p_\mu q^2-2q_\mu (q\cdot p)]\nonumber \\
 &&-\frac{1}{\pi}\int_{s_0}^{\infty} ds
\frac{\rho^h_T(s)[2p_\mu q^2-2q_\mu (q\cdot
p)]}{s-(p+q)^2}.\label{eq:spec2}
\end{eqnarray}
Here we have replaced the summations in (\ref{eq:sp1}) and
(\ref{eq:sp2}) with the dispersion integrations starting with the
threshold $s_0$ near the squared mass of the lowest scalar
$B$-meson\cite{Chernyak:1990ag}. The spectral densities can be
approximated as, by invoking the quark-hadron duality ansatz
\begin{eqnarray}
\rho^h_{+,-,T}(s)=\rho^{QCD}_{+,-,T}(s)\theta(s-s_0).
\end{eqnarray}
The QCD spectral densities $\rho^{QCD}_{+,-,T}(s)$ can be derived by
calculating the correctors in QCD theory. To this end, we work in
the large space-like momentum regions $(p+q)^2<<m^2_b$ for the
$b\bar{q_1}$ channel and a larger recoil region of the decaying
$B$-meson as given later, which correspond to the small light-cone
distance $x^2\approx0$ and are required by the validity of the OPE
\cite{Colangelo:2000dp}. Considering the effect of the background
gluon field, we can write down a full $b$-quark propagator
\begin{eqnarray}
&&\langle 0|Tb(x)\bar{b}(0)| 0\rangle \\
&&= i S_0(x,0) -ig_s\int \frac{d^4 k}{(2\pi)^4}e^{-ikx} \int
dv[\frac{\not\!k+m_b}{(m_b^2-k^2)^2}G^{\mu\nu}(vx)\sigma_{\mu\nu}+\frac{1}{m^2_b-k^2}vx_\mu
G^{\mu\nu}(vx)\gamma_\nu ].\nonumber
\end{eqnarray}
Here $G_{\mu\nu}$ is the gluonic field strength, $g_s$ denotes the
strong coupling constant and $S_0(x,0)$ expresses a free $b$-quark
propagator
\begin{eqnarray}
i S_0(x,0)=-i \int \frac{d^4
k}{(2\pi)^4}e^{-ikx}\frac{\not\!k+m_b}{m_b^2-k^2}.
\end{eqnarray}
The large virtuality of the underlying heavy quarks makes it sound
to neglect the contributions of soft gluon emission from the heavy
quarks, which, in fact, is just a twist-4 effect. In this accuracy
and leading order in $\alpha_s$, we find that as contrasted with the
results of the traditional LCSR \cite{Wang:2008da}, only the
nonlocal matrix element $\langle S(p)|\bar{q}_2(x)\gamma_\mu
q_1(0)|0 \rangle$ remains, while those concerning the nonlocal
operators $\bar{q}_2(x) q_1(0)$ and $\bar{q}_2(x) \sigma_{\mu\nu}
q_1(0)$ cancel out. As usual, applying the light-cone OPE to the
matrix element $\langle S(p)|\bar{q}_2(x)\gamma_\mu q_1(0)|0
\rangle$, we could be led to the leading twist-2 DA's of the scalar
mesons $\Phi_S(u,\mu)$ as defined in \cite{Cheng:2005nb}. We are
going to return to this point in the following section. Now the
light-cone OPE forms for the correlators can be written as follows:
\begin{eqnarray}
\Pi_{\mu}^{QCD}(p,q)&= &2ip_\mu m_b \int^1_0 du
\frac{\Phi_{S}(u)}{m^2_b-(q+up)^2},\\
\widetilde{\Pi}_{\mu}^{QCD}(p,q)&= &-2(p_\mu q^2-q_\mu (q\cdot p))
\int^1_0 du \frac{\Phi_{S}(u)}{m^2_b-(q+up)^2}.
\end{eqnarray}

We would like to convert them into a form of dispersion integration
in order to facilitate the ensuing subtraction of the effect of the
higher resonances and continuum states in the phenomenological
representations (\ref{eq:spec1}) and (\ref{eq:spec2}). To this end,
invoking the relation $m_b^2-(q+up)^2=u(s-(p+q)^2)$ we make a
replacement of $u$ with $s$. Matching both the forms of the
correlators, subtracting continuum contributions and making Borel
transformation \cite{Shifman:1978bx} with respect to the variable
$(p+q)^2$,
\begin{eqnarray}
B_{M^2}\frac{1}{m^2_B-(q+ p)^2}&=&\frac{1}{M^2}e^{-\frac{m^2_B}{M^2}},\nonumber\\
B_{M^2}\frac{1}{m^2_b-(q+up)^2}&=&\frac{1}{uM^2}e^{\frac{-1}{uM^2}\left[m^2_b+u(1-u)p^2-(1-u)q^2\right]},
\end{eqnarray}
with $M^2$ being the Borel parameter and $m_S$ the scalar meson
mass, we get the sum rules for the form factors:
\begin{eqnarray}
f_{+}(q^2) &=&-\frac{m_{q_1}+m_b}{m^2_B f_B} m_b
\int^1_{\Delta} du \frac{\Phi_S(u)}{u}e^{\Lambda} ,\label{eq:f+}\\
f_{-}(q^2) &=&\frac{m_{q_1}+m_b}{m^2_B f_B} m_b \int^1_{\Delta}
du \frac{\Phi_S(u)}{u}e^{\Lambda},\label{eq:f-}\\
f_T(q^2) &=&-\frac{m_{q_1}+m_b}{m^2_B f_B}(m_B+m_S) \int^1_{\Delta}
du \frac{\Phi_S(u)}{u}e^{\Lambda},\label{eq:ft}
\end{eqnarray}
where
\begin{eqnarray}
\Delta &=&\frac{1}{2m_S^2} \left[\sqrt{(s_0-m_S^2-q^2)^2 +
4(m_b^2-q^2) m_S^2}-(s_0-m_S^2-q^2)\right]
\label{eq:delta},\nonumber\\
\Lambda&=&-\frac{1}{uM^2}\left[m_b^2 +u(1-u)m_S^2-(1-u)q^2\right]
+\frac{m_B^2}{M^2} .
\end{eqnarray}
We find, as a by-product, that the form factors in question respect
the following LCSR relations:
\begin{eqnarray}
f_{+}(q^2)&=&-f_{-}(q^2),\label{eq:f-+}\\
f_T(q^2) &=& \frac{(m_B+m_S)}{m_b}f_{+}(q^2). \label{eq:ft+}
\end{eqnarray}

Actually, apart from that the same is observed in the LCSR involving
a pseudoscalar meson, a simple relation is obtained also for the
form factors in the vector meson case \cite{Huang:2008sn}. All these
observations, up to the hard-exchange corrections, are consistent
with the results of soft collinear effective theory (SCET)
\cite{Bauer:2000yr}. Having these relations at hand, in the
numerical discussion we will put our focus on the form factor
$f_+(q^2)$.

\section{Decay constants and distribution amplitudes of scalar mesons}

In this section, we give a brief review and discussion on the decay
constants and DA's of the related scalar mesons, which are the basic
inputs for the LCSR calculation.

For a light scalar meson in the two quark picture, it could couple
to the corresponding vector and scalar quark current operators thus
we can define its decay constants as \cite{Cheng:2005nb},
\begin{eqnarray}
\langle S(p)|\bar{q_2}(0)\gamma_\mu q_1(0)|0 \rangle &=& p_\mu f_S,\\
\langle S(p)|\bar{q_2}(0) q_1(0)|0 \rangle &=& m_S \bar{f}_S.
\end{eqnarray}
It is observed readily that the decay constants $f_S $ and
$\bar{f}_S$ are scale independent and dependent, respectively. The
neutral scalar mesons like $a^0_0$ and $f_0$( if considered purely a
$s\bar s$ bound state) cannot couple with a vector current operator
owing to the charge conjugation invariance or conservation of the
vector current and thus we have
\begin{eqnarray}
f_{f_0}=f_{a^0_0}=0.
\end{eqnarray}
For the other scalar mesons, the decay constants $f_S$ and
$\bar{f}_S$ are connected by equation of motion
\begin{eqnarray}
\bar{f}_S=\mu_S f_S, \label{eq:decay}
\end{eqnarray}
where
\begin{eqnarray}
\mu_S=\frac{m_S}{m_2(\mu)-m_1(\mu)},
\end{eqnarray}
the running quark masses $m_{i}(\mu)$ respect the renormalization
group equation (RGE):
\begin{eqnarray}
 m_i(\mu)&=& m_i(\mu_0)
  \left(\frac{\alpha_s(\mu_0)}{\alpha_s(\mu)}\right)^{-4/b},
\end{eqnarray}
with $b=(33 -2n_f)/3$, $n_f$ being the number of active quark
flavors. The decay constants $f_S$ hence are either zero or small of
order $ m_2-m_1$.

Similar to the case of pseudoscalar mesons, the twist-2 DA
$\Phi_S(u,\mu)$ of the scalar meson is defined as
\cite{Cheng:2005nb}
\begin{eqnarray}
\langle S(p)|\bar{q_2}(x)\gamma_\mu q_1(y)|0 \rangle &=& p_\mu
\int^1_0 due^{iup\cdot x+\bar{u}p\cdot y}\Phi_S(u,\mu),\label{eq:da}
\end{eqnarray}
with $u$ being the fraction of the light-cone momentum of the scalar
meson carried by $q_2$ and $\bar{u}=1-u$, and obeys the
normalization
\begin{eqnarray}
\int^1_0 du\Phi_S(u,\mu)&=&f_S\label{decay constant}.
\end{eqnarray}

With reference to the DA's of scalar mesons, a few words should be
given. From the definition of $\Phi_S(u,\mu)$, the corresponding
scalar mesons have to carry a large light-cone momentum $p_0+p_3$.
Along with the requirement of the OPE validity, such a constrain
condition demands that we work in a region assigned as,
\begin{eqnarray}
0\leq q^2<(m_b-m_S)^2-2(m_b-m_S)\Lambda_{QCD},
\end{eqnarray}
which, to be specific, is $0\leq q^2<11 \mbox{GeV}^2$ for a scalar
meson below $1~\mbox{GeV}$ and $0\leq q^2<8 \mbox{GeV}^2$ for one
above $1~\mbox{GeV}$. Also, it is important to realize that the DA's
of scalar meson, strictly speaking, become meaningful just at a
scale $\mu\geq m_S$, since the constituent quark of the scalar meson
is in essence off-shell and in particular, it is far from its mass
shell by the virtuality of $m_S^2$ as carrying the total momentum of
the scalar meson. Considering the DA's at a scale below $m_S$ means
that we are dealing with the situation that these off-shell modes
are in part or in full integrated out, however, which is
meaningless.

Based on the conformal symmetry hidden in the QCD Lagrangian,
$\Phi_S(u,\mu)$ can be expanded in a series of Gegenbauer
polynomials $C^{3/2}_{m}(x)$ with increasing conformal spin as
\begin{eqnarray}
\Phi_S(u,\mu)&=&\bar{f}_S(\mu)6u\bar{u}\left[B_0(\mu)+\sum_{m=1}B_m(\mu)C^{3/2}_m
(2u-1)\right], \nonumber  \label{eq:ge}
\end{eqnarray}
where Gegenbauer moments $B_m(\mu)$, which are scale dependent, are
given as
\begin{eqnarray}
B_m(\mu)=\frac{1}{\bar
f_s}\frac{2(2m+1)}{3(m+1)(m+2)}\int_0^1C_m^{3/2}(2u-1)\Phi_S(u,\mu)du.
\end{eqnarray}
The scale evolutions of $\Phi_S(u,\mu)$ are determined using the
following RGE:
\begin{eqnarray}
\bar f_S (\mu)&=&\bar f_S(\mu_0)\Bigg(
\frac{\alpha_s(\mu_0)}{\alpha_s(\mu)}
\Bigg)^{4/b},\nonumber\\
B_m(\mu)&=& B_m(\mu_0)
  \left(\frac{\alpha_s(\mu_0)}{\alpha_s(\mu)}\right)^{-(\gamma_{(m)}+4)/{b}},
  \label{eq:Bmomentdar}
\end{eqnarray}
where the one-loop anomalous dimensions is \cite{Gross:1974}
  \begin{eqnarray}
  \gamma_{(m)}= C_F
  \left(1-\frac{2}{(m+1)(m+2)}+4 \sum_{j=2}^{m+1}
  \frac{1}{j}\right),\nonumber
  \end{eqnarray}
with $C_F=4/3$. The conservation of charge parity demands an
antisymmetric $\Phi_S(u,\mu)$ under the interchange
$u\leftrightarrow 1-u$, namely, $\Phi_S(u,\mu)=-\Phi_S(1-u,\mu)$,
for the neutral scalar mesons of a $q\bar q$ content. Accordingly,
for the scalar mesons $a^0_0$ and $f_0$ we could write down their
leading twist DA's as
\begin{eqnarray}
\Phi_S(u,\mu)&=&\bar{f}_S(\mu)6u\bar{u}\sum_{m=0}B_{2m+1}(\mu)C^{3/2}_{2m+1}
(2u-1).
\end{eqnarray}
In the two quark picture, it is concluded that the twist-2 DA's of
all the light scalar mesons are antisymmetric under the interchange
$u\leftrightarrow 1-u$ in the flavor $SU(3)$ limit, thus the odd
Gegenbauer moments dominate in the DA's, forming a striking contrast
to the corresponding situations of the pseudoscalar mesons where the
leading DA of the pion, for instance, covers no odd Gegenbauer
moments and so is symmetric. Indeed the zeroth Gegenbauer moment
$B_0$, which is equal to $\mu^{-1}_S$, vanishes in the $SU(3)$
limit. In the following, we will neglect the contributions of the
even Gegenbauer moments and take only into account the first two odd
moments.

\begin{figure}
\centering
\includegraphics[scale=0.70]{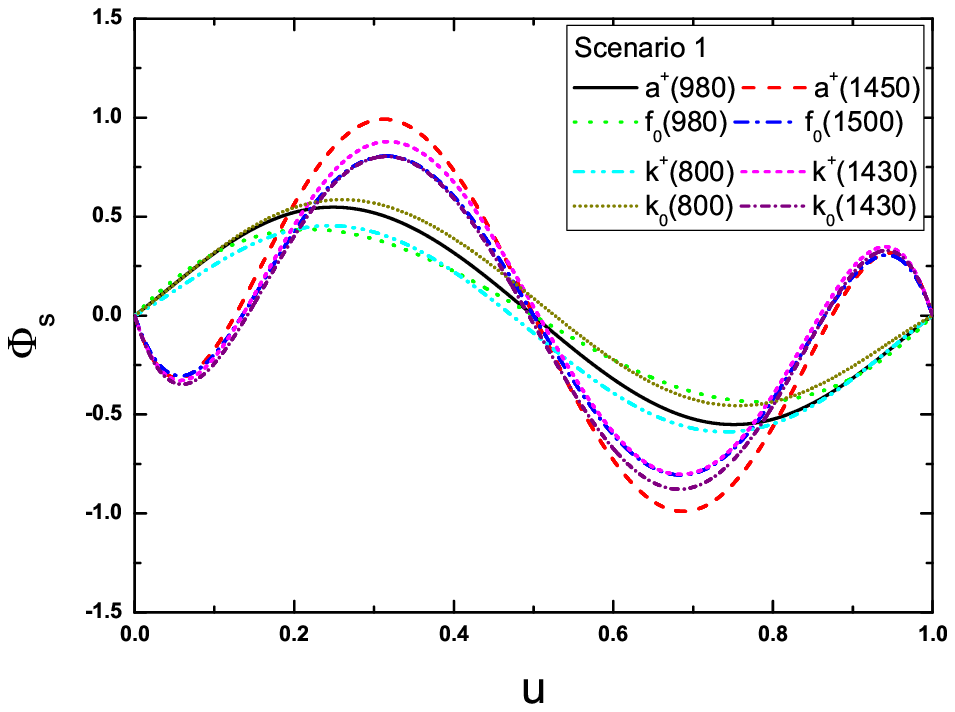}
\includegraphics[scale=0.70]{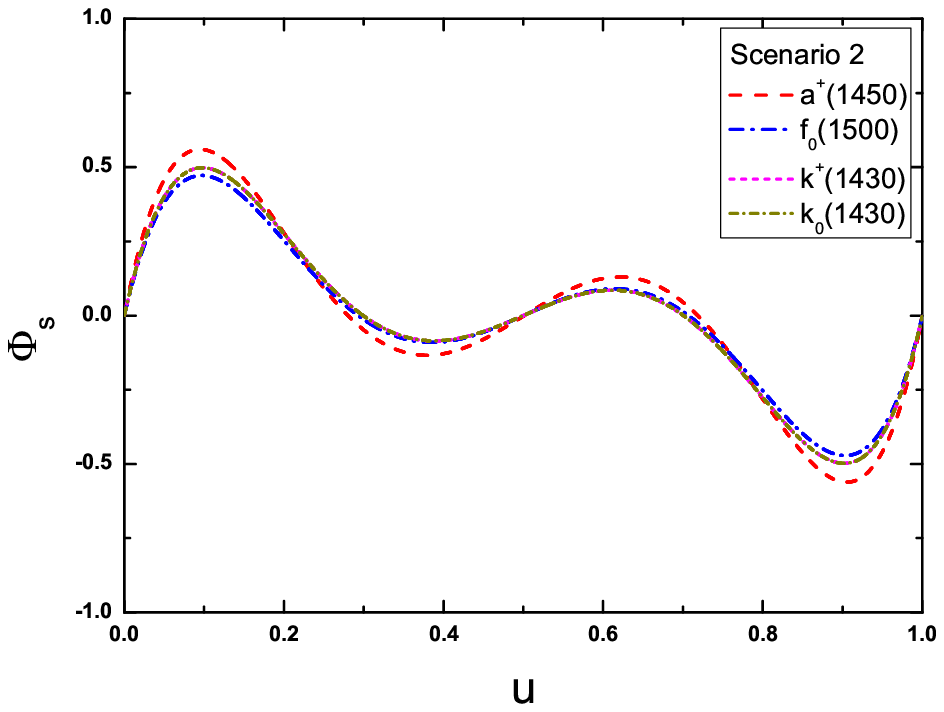}
\caption{Leading twist distribution amplitudes $\Phi_S$ of the
scalar mesons in scenario 1 and scenario 2 at the scale $\mu=2.4
~\mbox{GeV}$ . It can be seen that $\Phi_S$ is antisymmetric under
the replacement of $u\leftrightarrow1-u$ in the SU(3) limit owing to
the conservation of $C$ parity. }\label{da}
\end{figure}

To proceed, we must add that the LCSR for the form factor
$f_{+}(q^2)$ would have a distinct scale dependence, due to the
absence of the QCD radiative corrections. In such a case, it should
be in order that we work at the scale
$\mu_b=\sqrt{m_{B_s}^2-m_b^2}$, which denotes the typical virtuality
of the underlying $b$ quark. At this scale, the related parameters
can be evaluated making use of the RGE (\ref{eq:Bmomentdar}) with an
initial scale $\mu_0 \geq m_S$. As the initial conditions we prefer
using the QCD sum rule estimates at $\mu=1 ~\mbox{GeV}$
\cite{Cheng:2005nb}, which though is a bit inadequate for the
situation involving the scalar mesons above $1\mbox{GeV}$. The
numerical results for $\bar f_S (\mu)$ and $B_{1,3}$ are collected
in Tab.\ref{tab:bs1} and \ref{tab:bs2}, and the shapes of the DA's
in the two scenarios are illustrated in Fig.\ref{da}.
\begin{table}[htb]
\caption{ Decay constants $\bar{f}_s$ and Gegenbauer moments
$B_{1,3}$ of the twist-2 DA's $\Phi_S$ at the scales $\mu=1$ GeV
\protect\cite{Cheng:2005nb} and 2.4 GeV (shown in parentheses) in
scenario 1.} \label{tab:bs1}
\begin{ruledtabular}
\begin{tabular}{cccc}
 State & $\bar{f}$ ($GeV$)
 & $B_1$ & $B_3$  \\ \hline\hline
 $a_0(980)$ & 0.365(0.465)
  & -0.93$\pm$ 0.10 (-0.59 $\pm$0.07) & 0.14 $\pm$0.08(0.07 $\pm$0.04)      \\ \hline
 $a_0(1450)$& -0.280 (-0.357)
  & 0.89 0.20(0.56 $\pm$0.14) & -1.38 0.18 (-0.71 $\pm$0.11)  \\ \hline\hline
 $f_0(980)$ & 0.370(0.472)
  & -0.78 0.08 (-0.49 $\pm$0.06) & 0.02 0.07(0.01 $\pm$0.04)      \\ \hline
 $f_0(1500)$ & -0.255 (-0.325)
  & 0.80 0.40(0.51 $\pm$0.28) & -1.32 0.14 (-0.68 $\pm$0.08)  \\ \hline\hline
 $\kappa(800)$ & 0.340(0.433)
  & -0.92 0.11 (-0.58 $\pm$0.08) & 0.15 0.09(0.08 $\pm$0.05)  \\ \hline
 $K_0^*(1430)$ & -0.300 (-0.382)
  & 0.58 0.07(0.37 $\pm$0.05) & -1.20 0.08 (-0.62 $\pm$0.05) \\
\end{tabular}
\end{ruledtabular}
\end{table}
\begin{table}[htb]
\caption{Decay constants $\bar{f}_s$ and Gegenbauer moments
$B_{1,3}$ of the twist-2 DA's $\Phi_S$ at the scales $\mu=1$ GeV
\protect\cite{Cheng:2005nb} and 2.4 GeV (shown in parentheses) in
scenario 2.} \label{tab:bs2}
\begin{ruledtabular}
\begin{tabular}{cccc}
 State & $\bar{f}$ ($GeV$)
 & $B_1$ & $B_3$  \\ \hline\hline
 $a_0(1450)$ & 0.460(0.586)
  & -0.58 0.12 (-0.37 $\pm$0.08) & -0.49 0.15 (-0.25 $\pm$0.09)  \\ \hline\hline
 $f_0(1500)$ & 0.490(0.625)
  & -0.48 0.11 (-0.30 $\pm$0.08) & -0.37 0.20 (-0.19 $\pm$0.12)  \\ \hline\hline
 $K_0^*(1430)$ & 0.445(0.567)
  & -0.57 0.13 (-0.36 $\pm$0.09) & -0.42 0.22 (-0.216 $\pm$0.13) \\
\end{tabular}
\end{ruledtabular}
\end{table}

\section{Numerical calculation and discussion}

We proceed to do the LCSR calculation in the two scenarios with the
scalar mesons in the two quark picture. For illustrative purpose it
is sufficient to take, as a case study, the processes: $\bar{B}^0
\to a^+_0(980)/a^+_0(1450)~l\bar{\nu}_l$, $\bar{B}^0_s \to
\kappa^{+}(800)/K_0^{*+}(1430)~l\bar{\nu}_l$, $\bar{B}^0 \to
\bar{\kappa}^0(800)/\bar{K}^*_0(1430)~l \bar{l}$ and $\bar{B}^0_s
\to f_0(980)/f_0(1500)~l \bar{l}$.

The following inputs
\cite{Nakamura:2010zzi,Wang:2008da,Khodjamirian:1998ji} will be
taken in the numerical analysis:

\begin{equation}
\begin{array}{ll}
G_F=  1.166 \times 10^{-2} {\rm{GeV}^{-2}}, &
|V_{ub}|=3.96^{+0.09}_{-0.09} \times 10^{-3},
\\
|V_{tb}|=0.9991,  & |V_{ts}|=41.61^{+0.10}_{-0.80} \times 10^{-3},
\\
m_u(1 {~\mbox{GeV}})=2.8 ~\mbox{MeV}, &  m_d(1 ~\mbox{GeV})=6.8
~\mbox{MeV},\\
 m_s(1 {~\mbox{GeV}})=142 ~\mbox{MeV}, &m_b=(4.8 \pm
0.1) ~\mbox{GeV},
\\
m_{e,\mu}=0 ~\mbox{MeV},&m_{\tau}=1776.82 ~\mbox{MeV} ,
\\
 m_{B_0}=5.279~\mbox{GeV}, & m_{B_s}=5.368 ~\mbox{GeV},
\\
f_{B_0}=(0.19 \pm 0.02) ~\mbox{GeV},&  f_{B_s}=(0.23 \pm 0.02)
~\mbox{GeV}. \label{inputs}
\end{array}
\end{equation}

\begin{table}[htb]
\caption{Form factors $f_+$ and $f_-$ at zero momentum transfer
$q^2=0 ~\mbox{GeV}^2$ in scenario 1(S1) and scenario 2(S2) for
semileptonic decays $B_{(s)} \to S l^- \bar{\nu}_l$ with light-cone
sum rules(LCSR)\protect\cite{Wang:2008da}, sum
rules(SR)\protect\cite{Yang:2005bv} and perturbative
QCD(pQCD)\protect\cite{Li:2008tk} approaches.} \label{tab:fq01}
\begin{tabular}{|c|c|c|c|c|c|c|c|c|}
\hline
&\multicolumn{2}{c}{$\bar{B}^0_s \to K^{*+}_0(1430)$}
&\multicolumn{2}{|c}{$\bar{B}^0 \to a^{+}_0(1450)$}
&\multicolumn{2}{|c}{$\bar{B}^0_s \to \kappa^{+}(800)$}
&\multicolumn{2}{|c|}{$\bar{B}^0 \to a^{+}_0(980)$}\\[5pt]\cline{2-9}
Methods
&$f_+$&$f_-$&$f_+$&$f_-$&$f_+$&$f_-$&$f_+$&$f_-$\\
\hline
This work(S1)
 &$+0.10$&$-0.10$&$+0.26$&$-0.26$&$+0.53$&$-0.53$&$+0.56$&$-0.56$ \\ \hline
This work(S2)
 &$+0.44$&$-0.44$&$+0.53$&$-0.53$&$-$&$-$&$-$&$-$ \\ \hline
SR \protect\cite{Yang:2005bv}
 &$+0.24$&$-$&$-$&$-$&$-$&$-$&$-$&$-$ \\ \hline
LCSR(S2) \protect\cite{Wang:2008da}
 &$+0.42$&$-0.34$&$+0.52$&$-0.44$&$-$&$-$&$-$&$-$ \\ \hline
pQCD(S1) \protect\cite{Li:2008tk}
 &$-0.32$&$-$&$-0.31$&$-$&$+0.29$&$-$&$+0.39$&$-$ \\ \hline
pQCD(S2) \protect\cite{Li:2008tk}
 &$+0.56$&$-$&$+0.68$&$-$&$-$&$-$&$-$&$-$ \\ \hline
\end{tabular}
\end{table}

\begin{figure}
\centering
\includegraphics[scale=0.70]{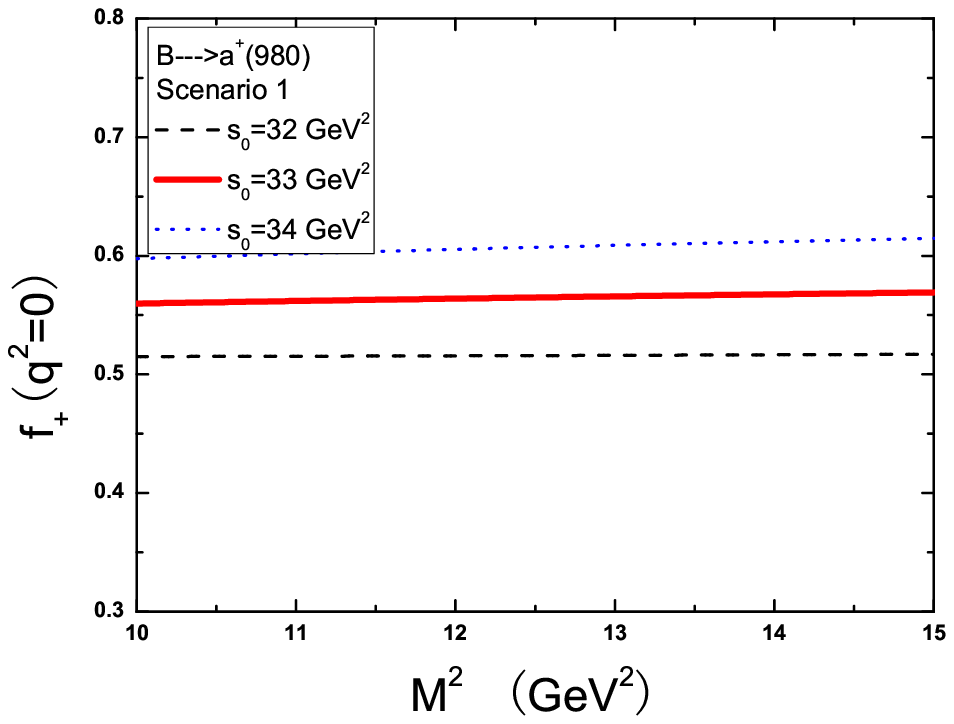}
\includegraphics[scale=0.70]{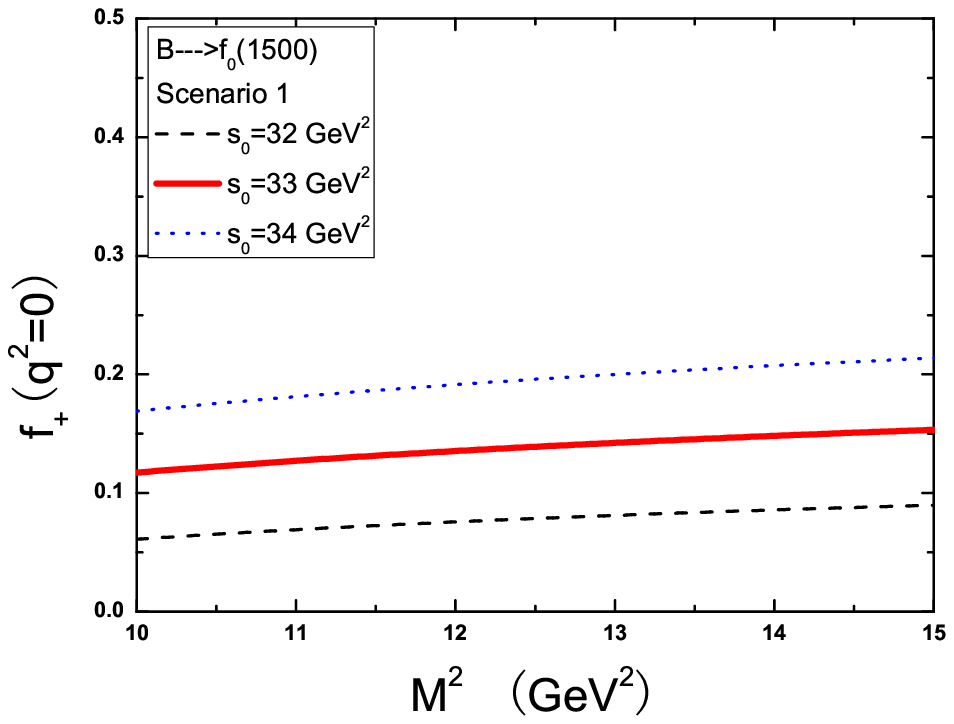}
\caption{Dependance of form factors $f_+(q^2=0)$ for $\bar{B}^0 \to
a^+_0(980)$ and $\bar{B}^0_s \to f_0(1500)$ on the Borel parameter
$M^2$ in scenario 1 within the LCSR approach at the scale $\mu=2.4
~\mbox{GeV}$. We take the threshold $s_0=32,33,34
~\mbox{GeV}^2$\cite{Chernyak:1990ag} and $b$ quark mass $m_b=4.8
~\mbox{GeV}$.}\label{fq0s1}
\end{figure}

\begin{table}[htb]
\caption{Form factors $f_+$, $f_-$ and $f_T$ for rare decays
$B_{(s)}\to S l \bar{l}$ at $q^2=0 ~\mbox{GeV}^2$ in scenario 1 (S1)
and scenario 2 (S2), with light cone sum
rules(LCSR)\protect\cite{Wang:2008da,Colangelo:2010bg}, sum
rules(SR)\protect\cite{Ghahramany:2009zz,Aliev:2007rq}, light front
quark model(LFQM)\protect\cite{Chen:2007na}, minimal supersymmetric
standard model(MSSM)\protect\cite{Aslam:2009cv}, covariant
light-front(CLF)\protect\cite{Cheng:2003sm}, covariant quark
model(CQM)\protect\cite{ElBennich:2008xy} and perturbative QCD
(pQCD)\protect\cite{Li:2008tk} approaches.} \label{tab:fq02}
\begin{tabular}{|c|c|c|c|c|c|c|c|c|c|c|c|c|}
\hline
&\multicolumn{3}{c}{$\bar{B}^0 \to \bar{K^*_0}(1430)$}
&\multicolumn{3}{|c}{$\bar{B}^0_s \to f_0(1500)$}
&\multicolumn{3}{|c}{$\bar{B}^0 \to \bar{\kappa}^0(800)$}
&\multicolumn{3}{|c|}{$\bar{B}^0_s \to f_0(980)$}\\[5pt]\cline{2-13}
Methods
&$f_+$&$f_-$&$f_T$&$f_+$&$f_-$&$f_T$&$f_+$&$f_-$&$f_T$&$f_+$&$f_-$&$f_T$\\
\hline
This work(S1)
 &$+0.17$&$-0.17$&$+0.24$&$+0.14$&$-0.14$&$+0.20 $&$+0.46$&$-0.46$&$+0.58$&$+0.44$&$-0.44$&$+0.58$ \\ \hline
This work(S2)
 &$+0.49$&$-0.49$&$+0.69 $&$+0.41$&$-0.41$&$+0.59$&$-$&$-$&$-$&$-$&$-$&$-$ \\ \hline
LFQM \protect\cite{Chen:2007na}
  &$-0.26$&$+0.21$&$-0.34$&$-$&$-$&$-$&$-$&$-$&$-$&$-$&$-$&$-$\\ \hline
CLF \protect\cite{Cheng:2003sm}
 &$+0.26$&$-$&$-$&$-$&$-$&$-$&$-$&$-$&$-$&$-$&$-$&$-$ \\ \hline
SR(S2) \protect\cite{Aliev:2007rq}
  &$+0.31$&$-0.31$&$-0.26$&$-$&$-$&$-$&$-$&$-$&$-$&$-$&$-$&$-$ \\ \hline
SR \protect\cite{Ghahramany:2009zz}
   &$-$&$-$&$-$&$-$&$-$&$-$&$-$&$-$&$-$&$+0.12$&$-0.17$&$-0.08$ \\ \hline
LCSR(S2) \protect\cite{Wang:2008da}
 &$+0.49$&$-0.41$&$+0.60$&$+0.43$&$-0.37$&$+0.56 $&$-$&$-$&$-$&$-$&$-$&$-$ \\ \hline
LCSR \protect\cite{Colangelo:2010bg}
  &$-$&$-$&$-$&$-$&$-$&$-$&$-$&$-$&$-$&$+0.19$&$-$&$+0.23$ \\ \hline
pQCD(S1) \protect\cite{Li:2008tk}
 &$-0.34$&$-$&$-0.44$&$-0.26$&$-$&$-0.34$&$+0.27$&$-$&$+0.29$&$+0.35$&$-$&$+0.40$ \\ \hline
pQCD(S2)\protect\cite{Li:2008tk}
  &$+0.60$&$-$&$+0.78$&$+0.60$&$-$&$+0.82$&$-$&$-$&$-$&$-$&$-$&$-$ \\ \hline
CQM \protect\cite{ElBennich:2008xy}
   &$-$&$-$&$-$&$-$&$-$&$-$&$+0.40$&$-$&$-$&$-$&$-$&$-$ \\ \hline
MSSM \protect\cite{Aslam:2009cv}
 &$+0.49$&$-0.41$&$+0.60$&$-$&$-$&$-$&$-$&$-$&$-$&$-$&$-$&$-$ \\ \hline
\end{tabular}
\end{table}

In the first place, let us make investigation in the context of
scenario 1. The numerical discussions of the form factors $f^+(q^2)$
can proceed in terms of the standard procedure for sum rule
calculations. The threshold parameters $s_0$, which correspond to
the masses $m^B_S $ of the lowest scalar $B_{(s)}$ mesons
\cite{Chernyak:1990ag}, need to be estimated in a certain
nonperturbative approach. Using the QCD sum rule result
\cite{Huang:1998sa} for the binding energy difference between the
scalar and pseudoscalar B mesons in the heavy quark effective
theory, we could give reasonably
$s^{\bar{B_0}}_0=s^{\bar{B^0_s}}_0=33 \pm 1~\mbox{GeV}^2$, which is
smaller than the threshold values in the corresponding conventional
sum rule calculations, with the experimental values of the
pseudoscalar B mesons. Also, it is possible to determine the
threshold parameters in other approaches, among which the scenario
suggested in \cite{Lucha:2009uy} is more effective. The range of the
Borel parameter $M^2$, which is shared by all the sum rules in
question, is determined as $10 ~\mbox{GeV}^2\leq M^2\leq 15
~\mbox{GeV}^2$. In this interval, the higher states and continuum
contribute less than $30\%$ and the sum rule results vary by $13\sim
30\%$ around the central values, depending on the decay modes.

\begin{figure}
\centering
\includegraphics[scale=0.70]{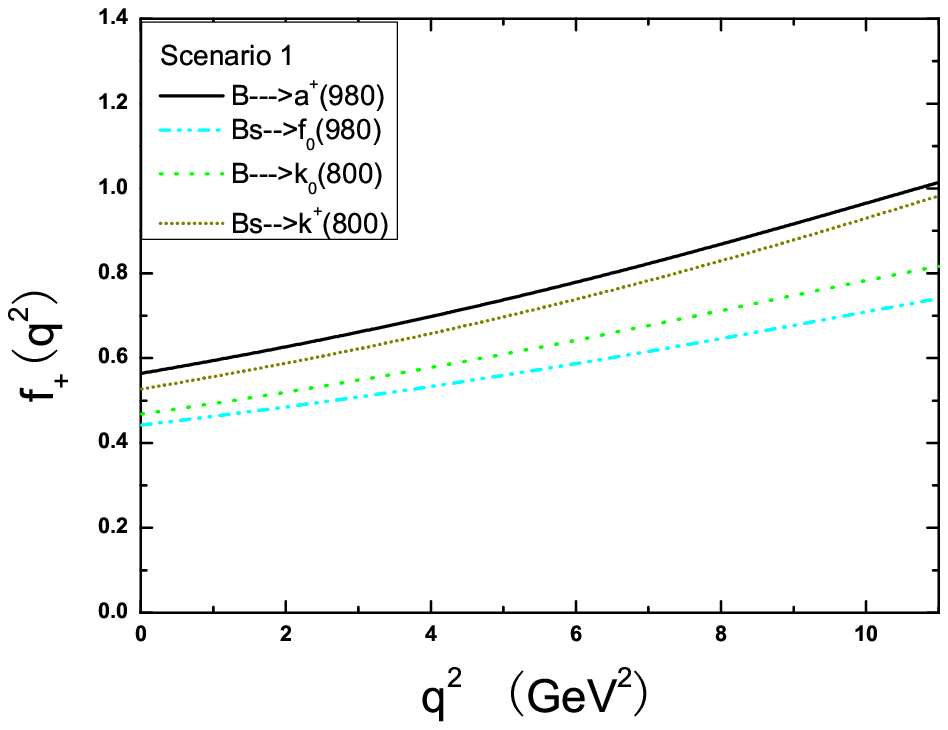}
\includegraphics[scale=0.70]{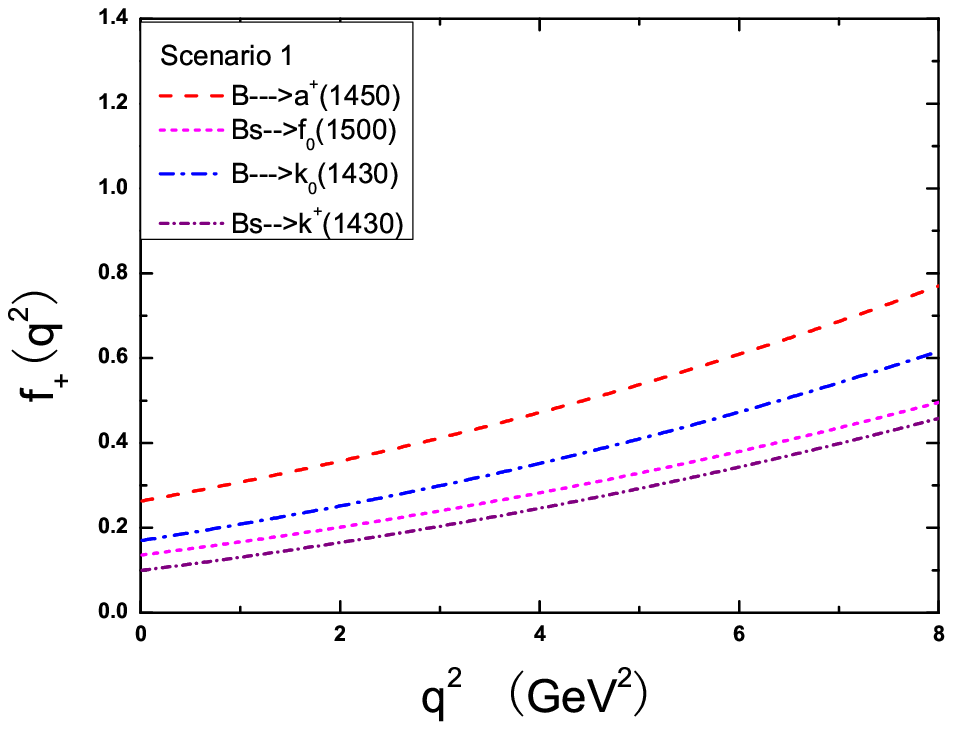}
\caption{Dependance of $B_{(s)} \to S$ form factors on the transfer
momentum $q^2$ in scenario 1 within the LCSR approach with the scale
$\mu=2.4 ~\mbox{GeV}$, threshold parameter $s_0=33 ~\mbox{GeV}^2$
and Borel parameter $M^2=12 ~\mbox{GeV}^2$.}\label{fqs1}
\end{figure}

To elucidate our findings for the form factors, we can consider
typically the case of the $B\to a_0(980)$ and $B_s\to f_0(1500)$
transitions. The LCSR for form factors, $f_+^{\bar{B^0}\to
a^+_0(980)}(0)$ and $f_+^{\bar{B^0_s}\to f_0(1500)}(0)$, are of a
good stability against $M^2$-varying, as shown in Fig.\ref{fq0s1}.
For simplicity, throughout the numerical investigation we give only
the central values of the sum rule results, corresponding to
$M^2=12\mbox{GeV}^2$ and $s^{\bar{B_0}}_0=s^{\bar{B^0_s}}_0=33
~\mbox{GeV}^2$. Then we have the observations $f_+^{\bar{B^0}\to
a^+_0(980)}(0)=0.56$ and $f_+^{\bar{B^0_s}\to f_0(1500)}(0)=0.14$.
Furthermore, use of the relations (\ref{eq:f-+}) and (\ref{eq:ft+})
leads to $f_-^{\bar{B^0}\to a^+_0(980)}(0)=-0.56$,
$f_-^{\bar{B^0_s}\to f_0(1500)}(0)=-0.14$ and $f_T^{\bar{B^0_s}\to
f_0(1500)}(0)=0.20$. Within the LCSR allowed kinematical regions,
$f_+^{\bar{B^0}\to ^+a_0(980)}(q^2)$ and $f_+^{\bar{B^0_s}\to
f_0(1500)}(q^2)$ as a function of $q^2$ are depicted in
Fig.\ref{fqs1}, along with those corresponding to the other modes.
The behaviors of $f_-^{B\to S}(q^2)$ and $f_T^{B\to S}(q^2)$ are
understandable likewise with the relations (\ref{eq:f-+}) and
(\ref{eq:ft+}). Additionally, for a complete understanding of
dynamical behaviors of the $B\to S$ transitions at the largest
recoils, one can be referred to Tab.\ref{tab:fq01} and
Tab.\ref{tab:fq02}, where we collect the present LCSR results for
the form factors $f_{+,-,T}^{B\to S}(0)$ in all the cases and the
predictions of other approaches for comparison.

\begin{figure}
\centering
\includegraphics[scale=0.70]{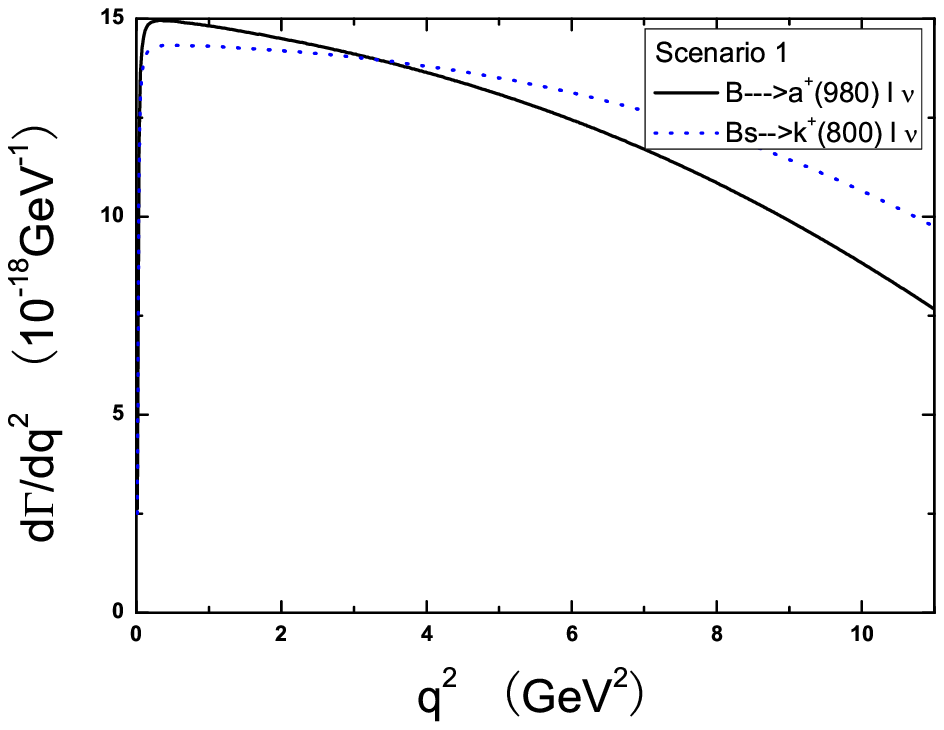}
\includegraphics[scale=0.70]{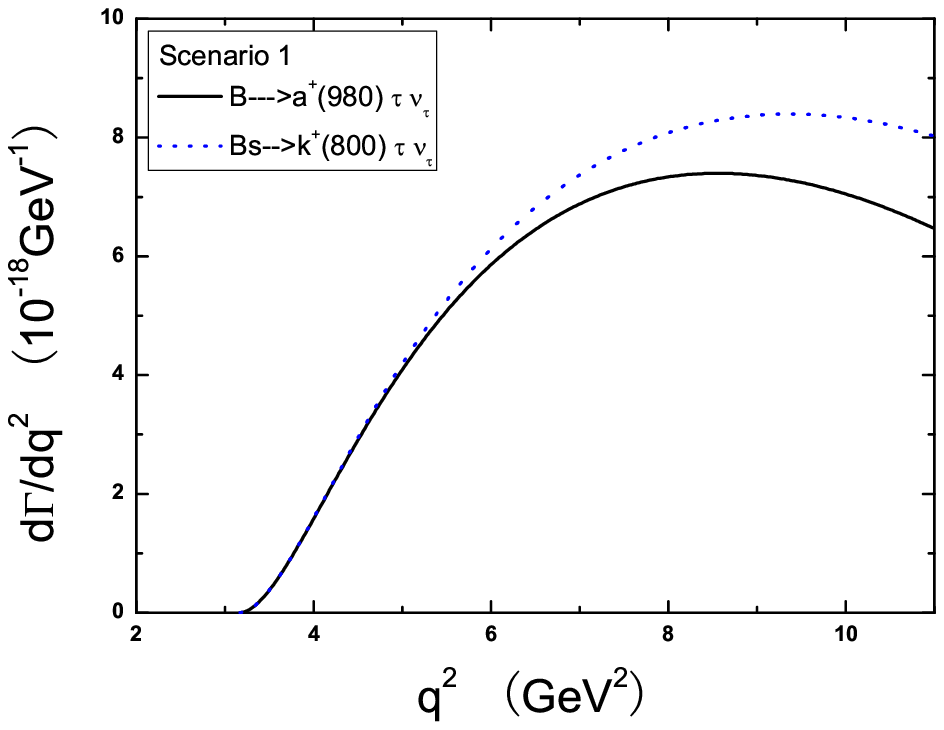}
\includegraphics[scale=0.70]{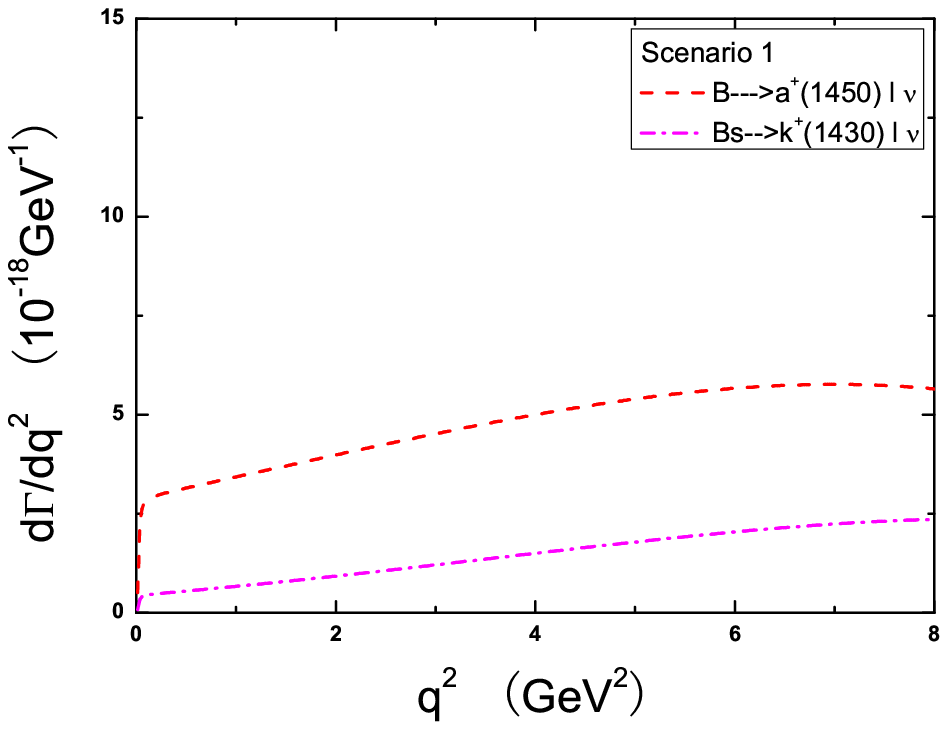}
\includegraphics[scale=0.70]{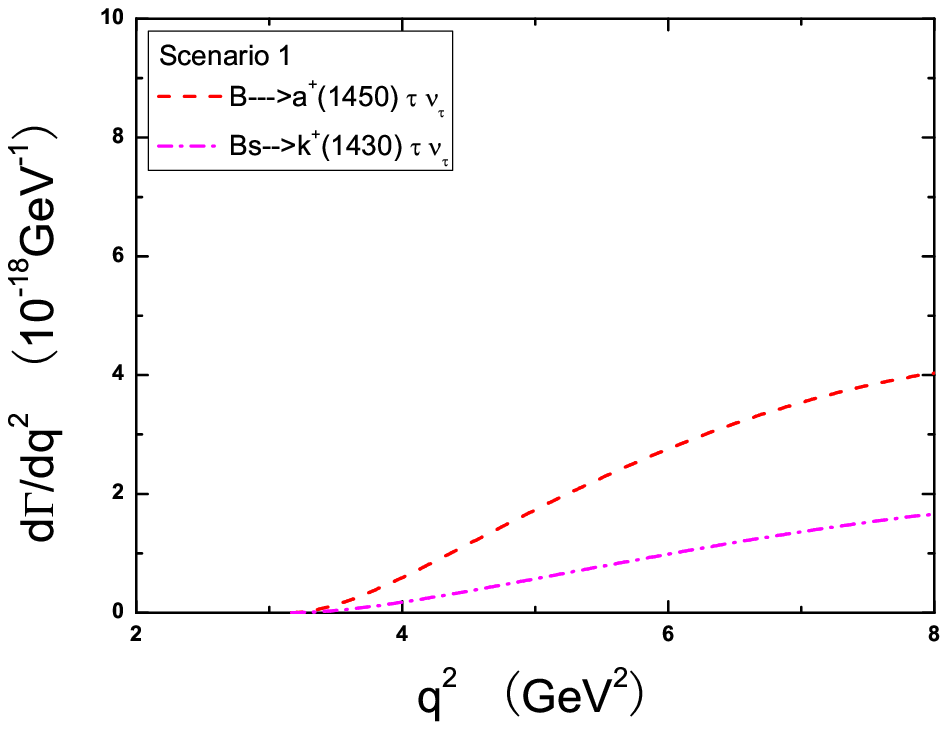}
\caption{Differential decay widths of the semileptonic $B\to
Sl\bar{\nu}_l$ decays as functions of $q^2$ in scenario 1. Here
$l=e,\mu$ in the left diagram.}\label{widthBlns1}
\end{figure}

\begin{figure}
\centering
\includegraphics[scale=0.70]{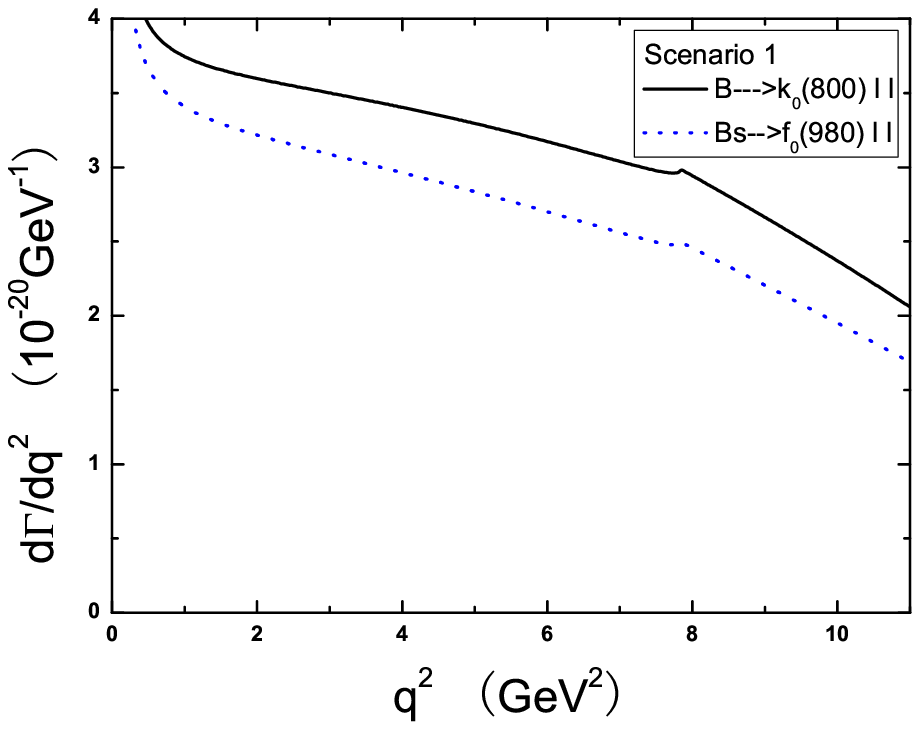}
\includegraphics[scale=0.70]{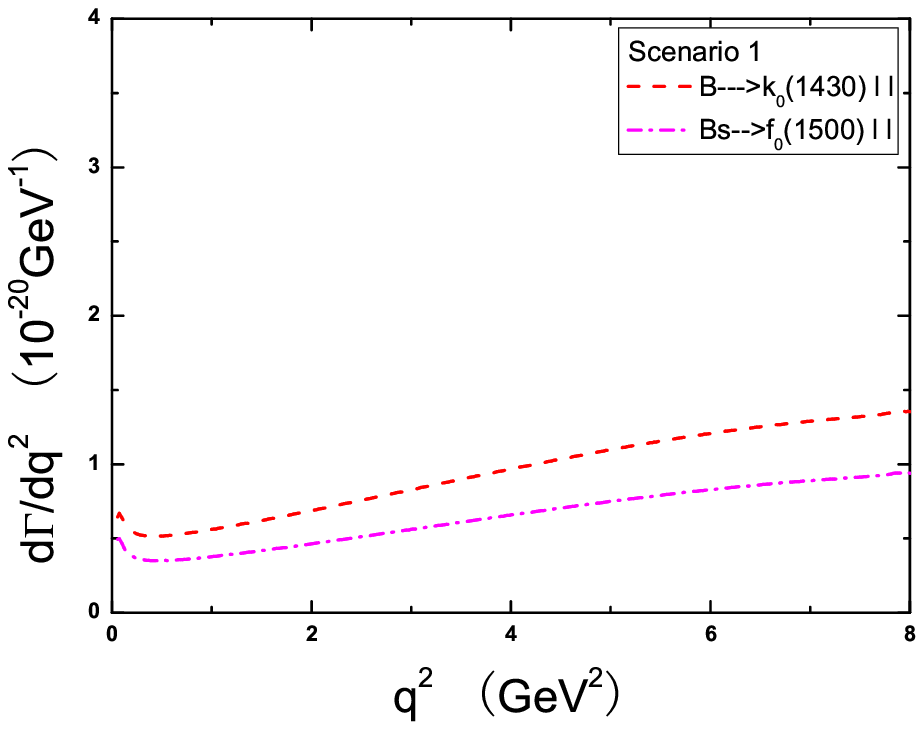}
\caption{Differential decay widths of the rare $B_{(s)}\to S l
\bar{l}$ ($l=e,\mu$) decays as functions of $q^2$ in scenario
1.}\label{widthBlls1}
\end{figure}

\begin{figure}
\centering
\includegraphics[scale=0.70]{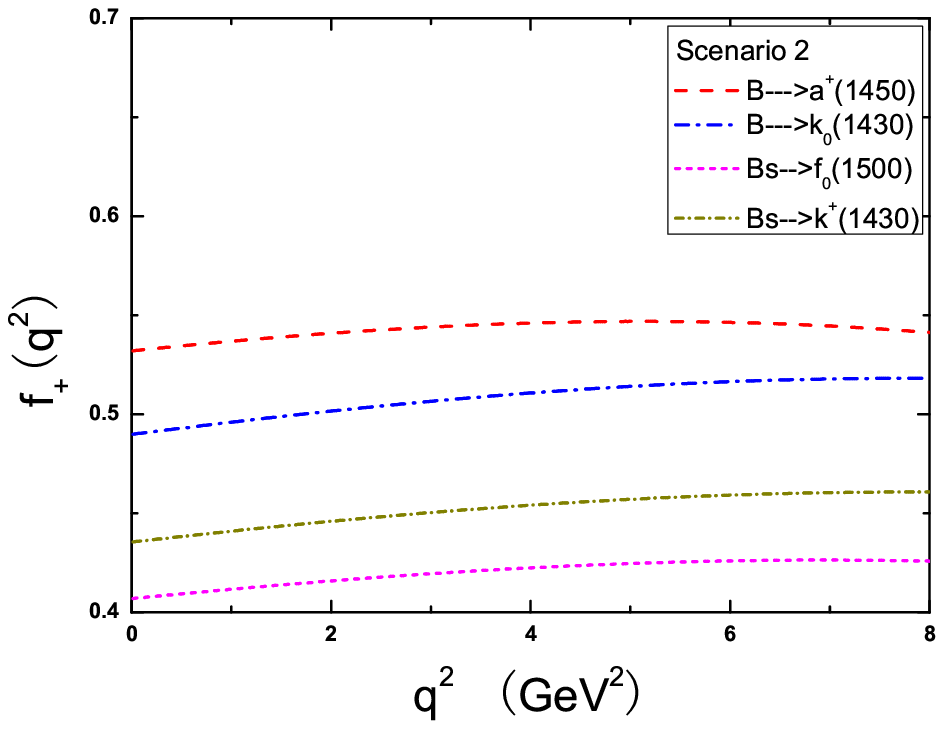}
\caption{Dependance of $B_{(s)} \to S$ form factors on the transfer
momentum $q^2$ in scenario 2 within the LCSR approach with the scale
$\mu=2.4 ~\mbox{GeV}$, threshold parameter $s_0=33 ~\mbox{GeV}^2$
and Borel parameter $M^2=12 ~\mbox{GeV}^2$.}\label{fqs2}
\end{figure}

\begin{figure}
\centering
\includegraphics[scale=0.70]{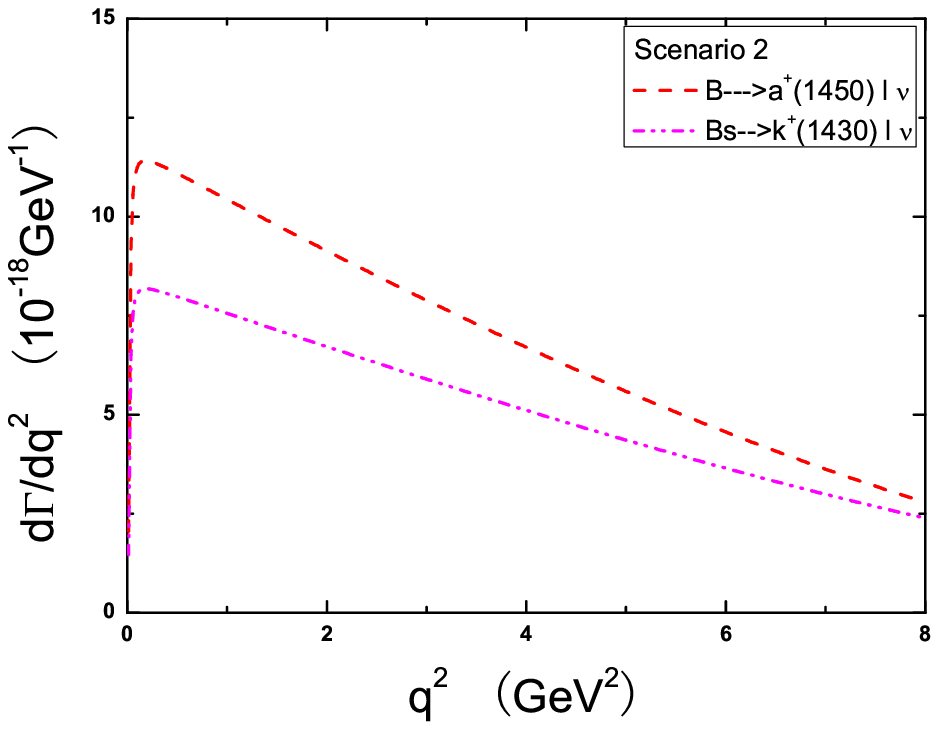}
\includegraphics[scale=0.70]{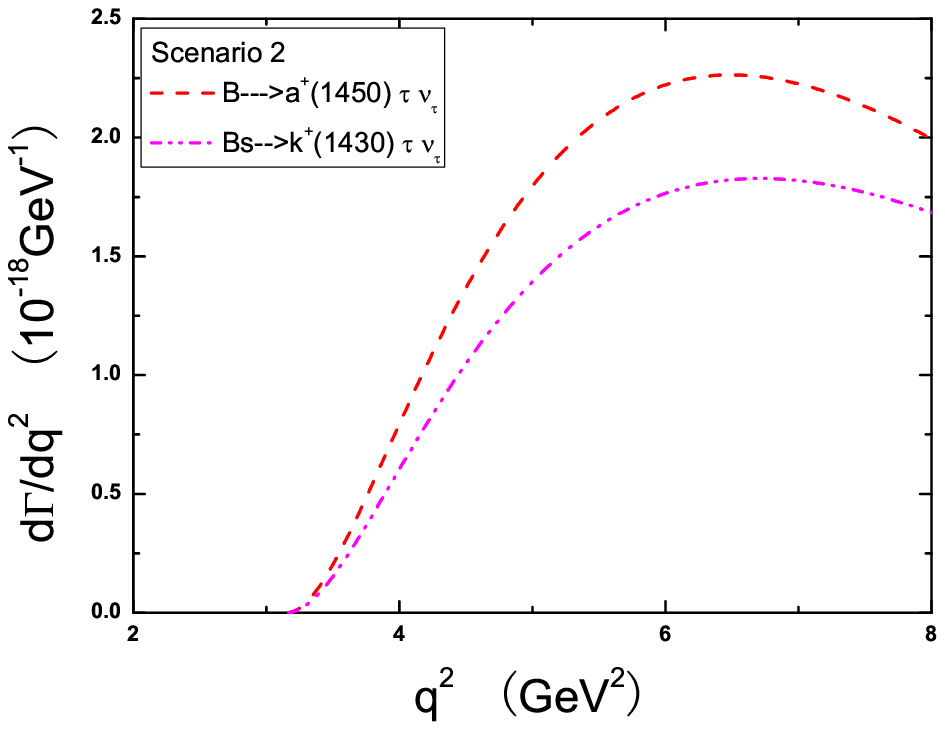}
\caption{Differential decay widths of the semileptonic $B\to
Sl\bar{\nu}_l$ decays as functions of $q^2$ in scenario 2. Here
$l=e,\mu$ in the left diagram.}\label{widthBlns2}
\end{figure}

\begin{figure}
\centering
\includegraphics[scale=0.70]{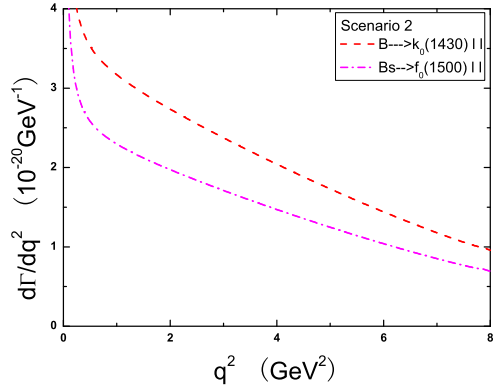}
\caption{Differential decay widths of the rare $B_{(s)}\to S l
\bar{l}$ ($l=e,\mu$) decays as functions of $q^2$ in scenario
2.}\label{widthBlls2}
\end{figure}

It is manifest that there is a sizable numerical difference in the
form factors between the transitions to the ground states and to the
excited ones. To make it clear, we go back to the LCSR expressions
for the form factors. We observe that the DA's $\Phi_S(u,\mu)$ make
contribution only in a smaller region of the momentum fraction $u$
ranging approximately from $0.8 \sim 1$ at $q^2=0$. The light quark
from the heavy quark decays prefers transferring to the region close
to its kinematical end-point to build a bound state with the
spectator quark of the decaying heavy meson, which is the so-called
Feynman mechanism, that is, soft-exchanges predominate over
hard-ones in the decay process. Referring to Fig.\ref{da}, one finds
that in that subregion the DA's behave quite differently between the
scalar objects below and above $1~\mbox{GeV}$. For the scalar mesons
below $1~\mbox{GeV}$, in the whole subregion their DA's turn out to
be negative and hence make a constructive contribution to the sum
rules. A different situation manifests itself as scalar mesons
involved are heavier ones: the DA's contribute constructively in one
part of the subrange but do destructively in the other. That the two
effects cancel out to a large degree leads to a form factor in
magnitude much smaller than those for the ground states. Physically,
this indicates that for a given $q^2$, as with the former situation
the decaying B mesons have a larger energy release in the latter
one.

In the same picture the $B\to S$ transitions have been explored in
the several approaches, such as the pQCD \cite{Li:2008tk}, QCD sum
rules \cite{Ghahramany:2009zz,Aliev:2007rq,Yang:2005bv} and LCSR
\cite{Wang:2008da,Colangelo:2010bg}. It is interesting to confront
our results with some of the previous studies. In what follows,
wherever a result of any other approach is referred, it should be
understood that we have, if necessary and possible, converted it
into that in the present convention. Application of the LCSR is
enforced to B decays to a scalar final state by taking the $B_s\to
f_0(980)$ semileptonic processes as a study case in
Ref.\cite{Colangelo:2010bg}. The sum rules for the form factors,
with the asymptotic forms used for twist-3 DA's, give
$f_+^{\bar{B^0_s}\to f_0(980)}(0)=0.19$ and $f_T^{\bar{B^0_s}\to
f_0(980)}(0)=0.23$ subject to an uncertainty estimate omitted here.
Counting QCD next-to-leading corrections, which is estimated roughly
based on the observation of the LCSR calculation for the $B\to \pi$
transitions, the above results are modified to $f^+_{B_s\to
f_0}(0)=0.24$ and $f^T_{B_s\to f_0}(0)=0.31$, about $45\%$ less than
our calculations. The reason for the sizable differences is mainly
use of the different inputs for the decay constant
$\bar{f}_{f_0(980)}$. The different scales are taken for the leading
and the subleading twist DA's as important inputs, which would have,
of course, an impact on accuracy of result. Using the same inputs,
the two evaluations are found to be consistent with each other and
that of QCD sum rules. The pQCD approach predicts, for the decay
modes to the scalar ground states, that the form factors are a bit
smaller in magnitude but within error comparable with the present
calculations, and have the approximately same value as in the case
of the first excited states, a result quite other than our
predictions. It is not difficult to understand for heavy-to-light
transitions, because the pQCD approach accords with the
hard-exchanges mechanism, and the resulting form factors rely on the
behaviors of the DA of light meson in the whole momentum region
accessible for the constitute quarks.

All the approaches mentioned above are no doubt applicable in the
kinematical region near the largest recoil for calculation of the
from factors. Nevertheless, no decisive region of $q^2$, in which
these approaches work well, has been provided in the existing
applications to the $B\to S$ transitions. In the LCSR calculation
\cite{Wang:2008da}, the form factors are artificially limited to the
range $0<q^2<15~\mbox{GeV}^2$, which seem somewhat large against our
estimate, and then the results are fitted to a dipole model for
having an understanding of the behaviors of the form factors in the
whole kinematically accessible region. The same way is adopted in
the pQCD calculation \cite{Li:2008tk} to extrapolate the results for
the form factors from small $q^2$ range to large one. Although such
a extrapolation manner is phenomenologically extensively assumed,
caution should be taken when one applies it to the present case.
First of all, we have no theoretical justification for doing so. The
pole models are believed to be suitable merely for description of
those form factors corresponding to $q^2$ near the squared pole
masses $m_{pole}^2$, however, for the present $B\to S$ transitions
the $m_{pole}^2$ are far away from their kinematical regions. On the
other hand, if the work region for an approach can not be assigned
effectively, choosing different fitting regions would lead to
different results. Hence, it is questionable to use a pole
description to get an all-around understanding of $q^2$ dependence
of the form factors for $B\to S$ transitions. Taking this into
account, we prefer calculating in the effective regions rather than
in the whole kinematical range.

Now, we are in a position to look into the differential decay rates
for the $B\to S$ semileptonic decays, which are expressed as
 \begin{eqnarray}
 \frac{d\Gamma}{dq^2}(B_{(s)}\to S l \bar{\nu}_l)&=&\frac{G_F^2|V_{ub}|^2}{192 \pi^3
 m_B^3}\frac{q^2-m_l^2}{(q^2)^2}\sqrt{\frac{(q^2-m_l^2)^2}{q^2}}
 \sqrt{\frac{(m_B^2-m_S^2-q^2)^2}{4q^2}-m_S^2}\nonumber\\
 &&\times\bigg[(m_l^2+2q^2)(q^2-(m_B-m_S)^2)(q^2-(m_B+m_S)^2)f_+^2(q^2)\nonumber\\
 &&~~~~+3m_l^2(m_B^2-m_S^2)^2\left(f_+(q^2)+\frac{q^2}{m_B^2-m_S^2}f_-(q^2)\right)^2\bigg],\label{eq:widthsemi}
 \end{eqnarray}
\begin{equation}
\frac{d\Gamma }{dq^2}(B_{(s)}\to S l \bar{l})=\frac{%
G_{F}^{2}\left| V_{tb}V_{ts}\right| ^{2}m_{B}^{3}\alpha _{em}^{2}}{%
1536 \pi ^{5}}\left( 1-\frac{4r_{l}}{s}\right) ^{1/2}%
\left[ \left( 1+\frac{2r_{l}}{s}\right)\varphi _{S}^{3/2} \alpha
_{S}+\varphi _{S}^{1/2}r_{l}\delta _{S}\right], \label{eq:widthrare}
\end{equation}%
where $m_l$ denotes the mass of a final state lepton, and
\begin{eqnarray*}
s &=&q^{2}/m_{B}^{2}\text{, \ \ \ \ \ \ \
}r_{l}=m_{l}^{2}/m_{B}^{2}\text{,
\ \ \ \ \ \ \ }r_{S}=m_{S}^{2}/m_{B}^{2}\text{,} \\
\varphi _{S} &=&\left( 1-r_{S}\right) ^{2}-2s\left( 1+r_{S}\right) +s^{2}%
\text{,} \\
\alpha _{S} &=& \left| C_{9}^{eff}{f_{+}\left( q^{2}\right) }
-2\frac{C_{7}f_{T}\left( q^{2}\right) }{1+\sqrt{r_{S}}}\right|
^{2}+\left|
C_{10}{f_{+}\left( q^{2}\right) }\right| ^{2} , \\
\delta _{S} &=&6\left| C_{10}\right| ^{2} \Big\{ \left[ 2\left(
1+r_{S}\right) -s\right] \left| {f_{+}\left( q^{2}\right)} \right|
^{2} + \left( 1-r_{S}\right) 2{\rm{Re}} [ f_{+}( q^{2})
f^{\ast}_{-}(q^2) ] +s\left| f_{-}(q^2)\right| ^{2}\Big\}.
\end{eqnarray*}

In the respective effective regions $m_l^2 \leq q^2\leq (m_B-m_S)^2$
and $4m_l^2 \leq q^2\leq (m_B-m_S)^2$, we assess the distributions
of the differential rates for the $B_{(s)}\to S l \bar{\nu}_l$ and
$B_{(s)}\to S l \bar{l}$, with the results displayed in
Fig.\ref{widthBlns1} and Fig.\ref{widthBlls1}, where we have set
$m_e=m_{\mu}=0$. In the case of $B_{(s)}\to S l \bar{l}$  there
appears a discontinuity at $q^2=4m_c^2$ stemming from the function
$h(z,s')$. The differential decay rates for the $B_{(s)}\to S
\tau^+\tau^-$ are incalculable in the present approach, for the
dilepton threshold $4m_{\tau}^2$ is beyond our work regions. It is
shown that our calculations and the predictions of pQCD
\cite{Li:2008tk} are comparable with each other, although they are
based on two different dynamical schemes.

As the scenario 2 is adopted, an analogous LCSR analysis can be made
in principle, however a complete discussion is not practicable at
present, due to little knowledge of the 4-quark scalar states below
$1~\mbox{GeV}$. Along the same line as above, we can assess the
semileptonic decays of $B_{(s)}$ to a scalar above $1~\mbox{GeV}$,
which is viewed as a two quark ground state. The sum rules show the
same Borel interval as in the case of scenario 1. The variations of
 the form factors $f^+(q^2)$ with $q^2$ are exhibited in
Fig.\ref{fqs2}, and at the largest recoil, a summary of the
numerical results for the form factors involved, including some of
the previous estimates, is given in Tab.\ref{tab:fq01} and
Tab.\ref{tab:fq02}. Comparing the sum rule calculations between the
scenarios 1 and 2, we see that in the the latter case the form
factors $f^+(q^2)$ have a central value between $0.40 \sim 0.70$ in
the effective regions, depending on the decay modes, and hence are
less sensitive to $q^2$ than in the former case in which there is a
large numerical range from $0.10 \sim 0.60$. The present evaluations
of $f^+(0)$, which show a better agreement with the conventional
LCSR calculation \cite{Wang:2008da,Colangelo:2010bg}, are a bit
smaller than the numerical observation in pQCD\cite{Li:2008tk}, and
meanwhile are large numerically in comparison with the calculation
of QCD sum rules in both the $B\to K^*$ and $B_{s}\to K^*$
situations, especially our result turning out to be about twice as
large as that of QCD sum rules in the latter case.

The resulting differential decay rates, as exhibited in
Fig.\ref{widthBlns2} and Fig.\ref{widthBlls2}, have a behavior other
significantly from what is observed in scenario 1, with the
remarkably different QCD dynamics embedded in the form factors
between the two scenarios. Once these scalar mesons above
$1~\mbox{GeV}$ are clearly identified to be, purely or mainly, the
two quark bound state, this result might help to distinguish between
both the pictures for them, as the future experiments become
accessible. In addition, the distribution shapes, which are
demonstrated by the differential rates for $B_{(s)}\to S l
\bar{\nu}_l$ in Fig.\ref{widthBlns2}, are compatible with the LCSR
calculation.

The decays to the scalar meson below $1~\mbox{GeV}$, despite
theoretically little accessible for the moment, could be discussed
qualitatively. In the four quark final states there is a
quark-antiquark component from the annihilations of emitted gluons
in the decaying processes, which gets the transitions highly
suppressed. Consequently we may deduce that in scenario 2 the
related form factors are of a small numerical value with respect to
the results in the two quark picture.

Finally, we should point out that all the above discussions can not
be generalized to $D_{(s)}$ decays to a scalar meson, because of the
fact that the decaying mesons have a recoil energy not large enough
to make LCSR applicable, in their decaying processes.

\section{Summary}

We have presented a LCSR computation on $B_{(s)}\to S l \bar{\nu}_l,
S l \bar{l}$ at leading order in $\alpha_s$, in the two quark
picture for the scalar mesons with the two different scenarios. A
correlation function with chiral current operator is chosen such
that the resulting LCSR the form factors can avoid the pollution
with the twist-3 DA's of the scalar mesons. Applicable regions of
the LCSR approach are discussed and are assigned reasonably as
$0\leq q^2 < 11 ~\mbox{GeV}^2$ and $0\leq q^2 < 8~\mbox{GeV}^2$, for
the scalar final states below and above $1~\mbox{GeV}$,
respectively. Also, we investigate the properties of the DA's of the
scalar mesons, obtaining an observable difference from the case of
the pseudoscalar mesons. In the effective regions, the form factors
and differential decay rates are estimated, with the main findings
summarized as the following: (1) There exist the relations among the
the form factors for the $B\to S$ transitions, which are in
accordance with the prediction of SCET. (2) For the decays to a
scalar ground state, in the case of scenario 1 the form factors at
$q^2=0$ show the numerical result much larger those for the first
excited sates, and as confronted with the corresponding observations
in scenario 2, the former seem large in magnitude, but the latter
are predicted to be small. (3) For the semileptonic processes with
the scalar final state above $1 ~\mbox{GeV}^2$, the resulting
differential decay rates have a significantly different behavior for
the different scenarios. Some of them might be beneficial to
experimentally identify physical natures of the scalar mesons. The
present results might be improved as the QCD radiative corrections
are taken into account, and however they are not expected to change
too much from the LCSR calculation on the $B\to \pi$ transition\cite{Wan:2002hz}.

\hspace{1cm}

{\bf Acknowledgments}: Y. J. Sun would like to thank Dr.Yu-Ming Wang
for helpful discussions. This work is supported by Natural Science
Foundation of China under Grant Nos.10735080, 10805082 and 10675098.

\end{document}